\documentclass[sigconf]{acmart}

\usepackage{enumitem}
\usepackage{amsmath}
\usepackage{amsfonts}
\usepackage{multirow}
\usepackage{threeparttable}
\usepackage{arydshln}
\usepackage{adjustbox}
\usepackage{graphicx}
\usepackage{algorithm}
\usepackage{algpseudocode}
\usepackage{listings}
\usepackage{xcolor}
\usepackage{bbding}
\usepackage{pifont}
\usepackage{balance}

\AtBeginDocument{%
  }

\copyrightyear{2026}
\acmYear{2026}
\setcopyright{cc}
\setcctype{by}
\acmConference[KDD 2026] {Proceedings of the 32nd ACM SIGKDD Conference on Knowledge Discovery and Data Mining V.2}{August 9--13, 2026}{Jeju Island, Republic of Korea.}
\acmBooktitle{Proceedings of the 32nd ACM SIGKDD Conference on Knowledge Discovery and Data Mining V.2 (KDD 2026), August 9--13, 2026, Jeju Island, Republic of Korea}
\acmISBN{979-8-4007-2259-2/2026/08}
\acmDOI{10.1145/3770855.3818362}

\begin{document}

\title{DEGR: Dual Exploration-Driven Generative Re-Ranking for Adaptive Cross-Request Context Bridging}


\author{Binglei Zhao}
\email{zhaobinglei1@jd.com}
% \authornote{Corresponding author.}
\orcid{0000-0002-7334-9452}
\affiliation{%
    \institution{JD.com}
    % \institution{Marketing \& Commercialization Center, JD.com}
    \city{Beijing}
    \country{China}
}

% \author{Feng Mei}
% \email{meifeng6@jd.com}
% \orcid{0009-0002-9442-2548}
\author{Xuanhua Yang}
\email{yangxuanhua1@jd.com}
\orcid{0009-0005-6644-627X}
\affiliation{%
    \institution{JD.com}
    % \institution{Marketing \& Commercialization Center, JD.com}
    \city{Beijing}
    \country{China}
}

% \author{Feng Mei}
% \email{meifeng6@jd.com}
% \orcid{0009-0002-9442-2548}
\author{Xiwei Zhao}
\email{zhaoxiwei@jd.com}
\orcid{0000-0002-9382-6041}
% \author{Sulong Xu}
% \email{xusulong@jd.com}
% \orcid{0000-0003-0345-334X}
\affiliation{%
    \institution{JD.com}
    % \institution{Marketing \& Commercialization Center, JD.com}
    \city{Beijing}
    \country{China}
}

\author{Sulong Xu}
\email{xusulong@jd.com}
\orcid{0000-0003-0345-334X}
\affiliation{%
    \institution{JD.com}
    % \institution{Marketing \& Commercialization Center, JD.com}
    \city{Beijing}
    \country{China}
}

\renewcommand{\shortauthors}{Binglei Zhao, Xuanhua Yang, Xiwei Zhao, and Sulong Xu.}

\begin{abstract}
In industrial recommendation systems, the re-ranking stage balances business objectives and diversity for sequence-level optimization while modeling contextual information.
However, constrained by fixed upstream supply, existing methods fail to deliver further effectiveness gains, especially under low-quality supply. 
To overcome this, re-ranking can actively balance immediate and exploratory value, for instance, by prioritizing exploratory exposure under low-quality supply to preserve browsing potential and facilitate serendipitous conversions.
Therefore, we propose a \textbf{Dual Exploration-Driven Generative Re-Ranking (DEGR)} method.
DEGR adopts a hybrid supervised-reinforcement \textbf{exploration and optimization} paradigm, guided by an \textbf{exploratory reward} model that adaptively balances immediate and exploratory value.
The hybrid optimization paradigm integrates three key components: supervised learning, exploration diversity constraint, and adaptive reward-weighted ORPO for preference optimization.
Through this \textbf{dual exploration}, the generator ultimately acts as an \textbf{adaptive cross-request contextual bridge}.
Offline and online experiments indicate that DEGR outperforms SOTA methods, achieving improvements of up to \textbf{1.22\%} UCTR and \textbf{0.20\%} PV in the JD E-commerce recommendation system.
\end{abstract}




% \received[accepted]{1 June 2026}

\maketitle

\begin{figure}
  \centering
  \includegraphics[width=0.48\textwidth, height=3.8cm]{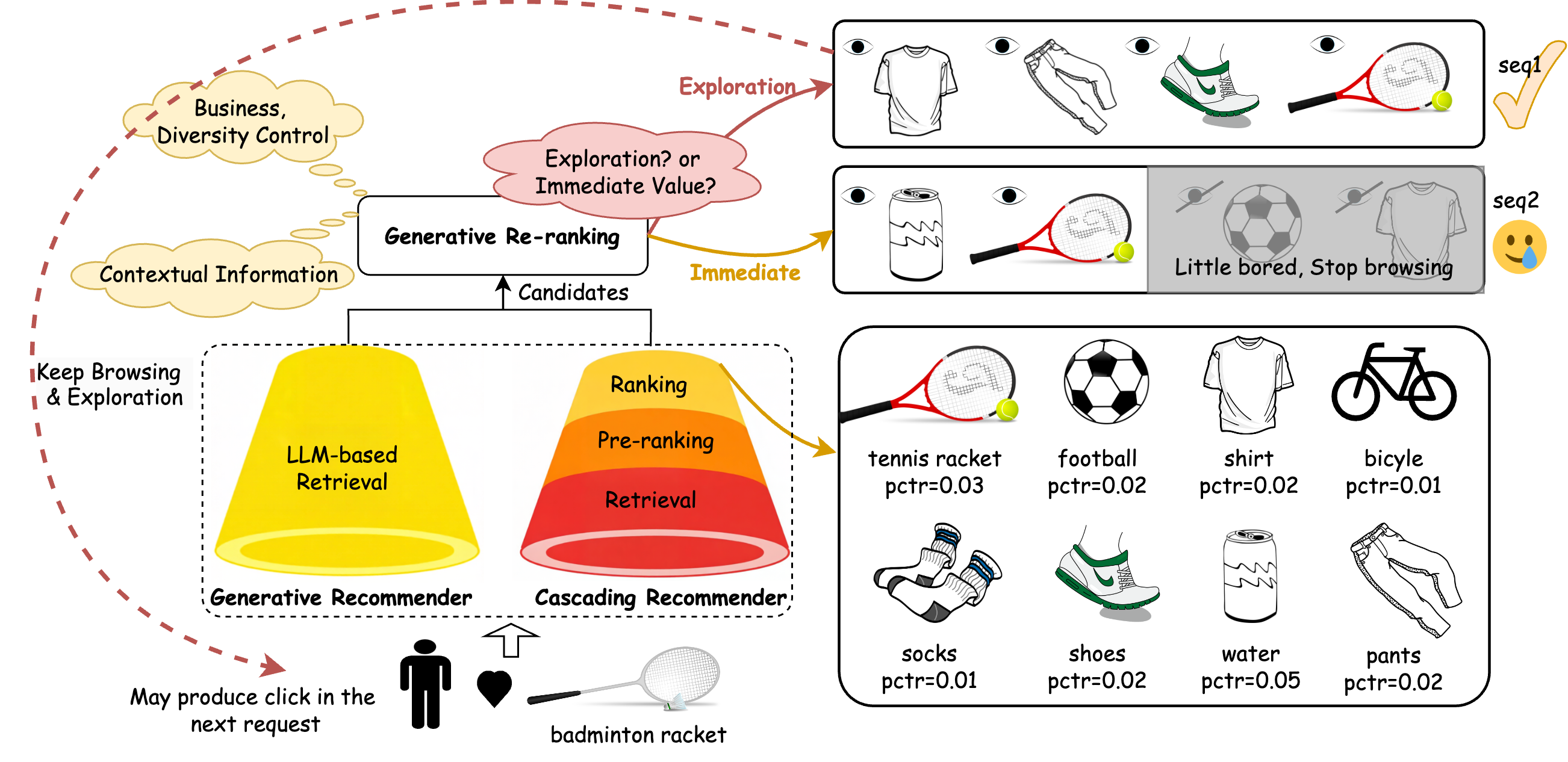}
  \caption{An example of different re-ranking sequences under the low-quality upstream supply. 
% After buying a badminton racket, the user receives fewer relevant candidates during the re-ranking stage.
Placing low-pCTR items first maintains browsing momentum, potentially leading to later conversions (Seq1). 
But putting high-pCTR water first, while initially effective, quickly bores users (Seq2).
Context affects user behavior toward the same item across varying sequence compositions, such as the tennis racket.
}
\label{fig:brief_next}
\end{figure}

\section{Introduction}
While modern industrial recommendation systems adopt novel architectures, from traditional cascading frameworks \cite{2020COLD, 2024prerank} to recent generative LLM-based paradigms \cite{2025onerec, 2025onerecv2}, re-ranking, the stage closest to the users, plays an important role in achieving business goals and controlling diversity, as illustrated in Figure\ref{fig:brief_next}. 
From the perspective of model structure, compared to ranking, which models item-wise click-through rate (CTR) or conversion rate (CVR), re-ranking needs to explore the optimal context-aware sequence among the vast space of permutations. 
From the optimization objectives, re-ranking requires maximizing sequence value to balance business goals and diversity, in addition to item-wise accuracy.

Re-ranking methods are typically categorized into one-stage, two-stage, and generator-only approaches.
One-stage methods are exemplified by PRM\cite{2019prm}, which refines scores by modeling contextual information and applies greedy search based on refined scores. 
However, since re-ranking stages reorder items, the final sequence constitutes an entirely new information environment, so the contextual dependencies learned become inconsistent with the actual exposure contexts the user perceives.
Two-stage methods propose the Generator-Evaluator (GE) framework, where the generator samples multiple sequences, and then the evaluator selects the optimal sequence by modeling actual exposure contexts.
Given the infeasibility of exhaustive permutation evaluation, recent two-stage methods prioritize efficient modeling of sequence distributions.
% Most of them utilize reward signals from the evaluator to guide generator's exploration, based on actor-critic (AC) reinforcement learning\cite{2015High}.
Then, some generator-only methods\cite{2025Gref} are proposed to generate a single sequence directly for online serving, guided by user feedback or reward models.

As the terminal stage shaping the final exposure outcomes in recommendation systems, re-ranking still faces inherent \textbf{upstream limitation}.
Given a fixed item supply from upstream stages, the solution space of re-ranking is inherently constrained, particularly under low-quality supply conditions.
% Existing methods do not consider the dynamic re-formulation of value, limiting re-ranking's potential.
This naturally raises a fundamental question: how to quantify the sequence value to amplify re-ranking effectiveness?
Thus, re-ranking can actively balance immediate and exploratory value, for instance, by prioritizing exploratory exposure under low-quality supply to facilitate latent conversions through deeper browsing, as shown in Figure\ref{fig:brief_next}.

\begin{figure}
  \centering
  \includegraphics[width=0.48\textwidth, height=1.9cm]{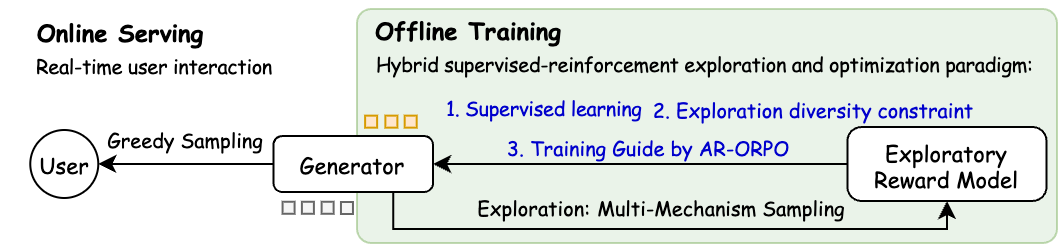}
  \caption{Offline training and online serving processes of our dual exploration-driven generative re-ranking.
}
\label{fig:all}
\end{figure}

To this end, we propose the \textbf{Dual Exploratory-Driven Generative Re-Ranking (DEGR)} method to balance immediate and exploratory value, further building an adaptive cross-request contextual bridge.
DEGR adopts a hybrid supervised-reinforcement \textbf{exploration and optimization} paradigm, guided by an \textbf{exploratory reward model}, as illustrated in Figure\ref{fig:all}.
% ---- reward model -----
The \textit{exploratory reward model} captures authentic contextual awareness and adaptively balances immediate and exploratory value.
It proactively preserves browsing potential and enables serendipitous conversions during exploration interactions. 
% ---- generator -----
The \textit{generator} employs an encoder-decoder architecture, using multiple decoding-head cohorts to generate sequences in parallel and efficiently.
% ---- optimization paradigm -----
The \textit{hybrid exploration and optimization paradigm} integrates three key components: 
% supervised learning, exploration diversity constraint, and adaptive reward-weighted ORPO for preference optimization. 
(1) \textbf{Supervised learning} aligns the generator with online distribution and avoids potential capability collapse in RL.
(2) \textbf{Exploration diversity constraint} achieves intra-cohort regularization when item parallel generation, effectively mitigating semantic redundancy in decoding cohorts.
(3) \textbf{Adaptive Reward-weighted ORPO} (AR-ORPO) guides the generator exploration to maximize exploratory rewards based on ORPO\cite{2024orpo}. 
It constructs preference lists and dynamically uses rewards as credible soft weights to refine the optimization trajectory.
% maximizes reward utility while suppressing noise.
To enhance exploration capacity, we introduce a \textbf{Multi-Mechanism Sampling Strategy} to yield diverse sequences.
Through the \textbf{dual exploration}, the generator acquires authentic contextual awareness, and ultimately acts as an adaptive cross-request contextual bridge.
% ---- online -----
When online serving, we only deployed the generator.
Contributions are summarized as follows:
% The generator ultimately masters generation capability as an adaptive "global sequence optimizer" across requests, rather than a dependent "local candidate proposer" relying on evaluators.
\begin{itemize} [leftmargin=*]
\item We propose the Dual Exploratory-Driven Generative Re-Ranking (DEGR) that adaptively balances exploratory value with immediate value to build a cross-request contextual bridge.
% It can be effectively deployed in industrial recommendation systems.

\item 
We propose the dual exploration, a hybrid exploration and optimization paradigm guided by an exploratory reward model, which helps to construct an optimal sequence generator.

\item Extensive experiments verify DEGR's effectiveness in offline datasets and online systems.
It outperforms state-of-the-art baselines, achieving a 1.22\% UCTR and 0.20\% PV in JD's recommendation system with only the generator deployed.
\end{itemize}

\section{Related Work}
% We briefly describe re-ranking methods and preference optimization below: 

\begin{figure*}
  \centering
  \includegraphics[width=1\textwidth, height=5.7cm]{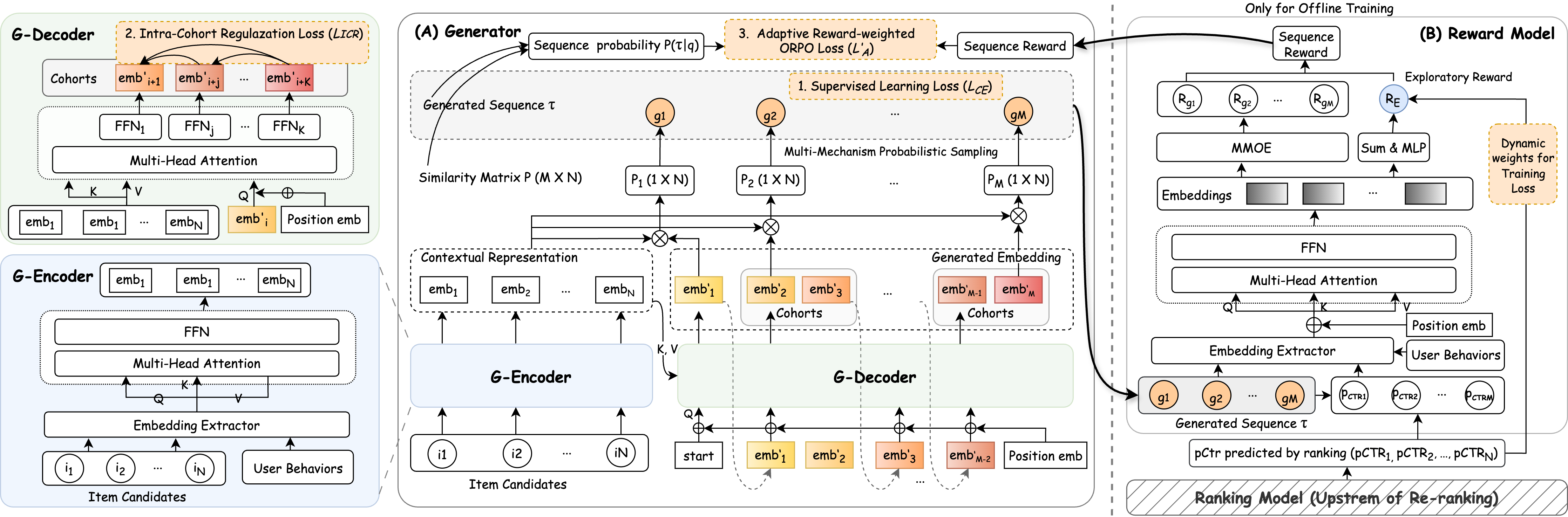}
  \caption{Architecture of Dual Exploratory-Driven Generative Re-ranking. (A) The encoder-decoder generator is trained using a hybrid exploration-and-optimization paradigm guided by an exploratory reward model.
  (B) The exploratory reward model uses pCTR predicted by upstream as dynamic weights to balance immediate and sequence-level exploratory rewards adaptively.
}
\label{fig:arch}
\end{figure*}

\subsection{Re-ranking}
In industrial recommendation systems, re-ranking aims to optimize sequence-level value, which is typically categorized into one-stage, two-stage, and generator-only approaches.
One-stage methods take ranking lists as inputs and model context information between items using RNN or self-attention to refine scores, such as DLCM\cite{2018DLCM}, MiRNN\cite{2018MiRNN}, PRM\cite{2019prm}, Seq2slate\cite{2019Seq2Slate}, Re-ranking in Kuaishou\cite{2022kuaishou}, MIR\cite{2022MIR}, and EXRT\cite{2022EXTR}.
Seq2slate\cite{2019Seq2Slate} compares the effectiveness of reinforcement and supervised learning.
However, they are constrained to point-wise score estimation, lacking the capacity to capture real exposure context, making them structurally incapable of achieving global optimal arrangements.
Two-stage GE (Generator-Evaluator) frameworks use generators to generate multiple sequences and evaluators to select the optimal sequence by modeling real-world exposure context.
And the generator is trained guided by evaluator based on actor-critic(AC) algorithms.
Recent two-stage methods prioritize efficient sequence candidate generation over traditional heuristic sampling due to the infeasibility of exhaustive permutation evaluation, as seen in PRS\cite{2021PRS} adopting beam-search, PIER\cite{2023PIER} applies SimHash, GRN\cite{2023GRN} using pointer network\cite{2017pointer}. 
Based on these, CMR\cite{2023CMR} proposes a multi-target combination of efficiency and business with a hypernetwork to achieve flexible control.
NAR4Rec\cite{2025NAR4Rec} proposes a non-autoregressive matching model to speed up sequence generation. 
To prevent local optima entrapment, MG-E\cite{2025MG_E} employs multiple generators to enhance sampling sequences' diversity and exploration.
Then, some generator-only methods\cite{2025Gref} are proposed to generate a single sequence directly for online serving, guided by user feedback or reward models.
Under the upstream limitation, these methods make re-ranking a passive corrective module; however, re-ranking inherently requires linking continuous requests, which is neglected in the above methods.

\subsection{Reinforcement Learning in Re-ranking}
% \subsection{Preference Optimization}
Actor-critic (AC) algorithms construct an actor for policy generation and a critic as a value function, and maximize the expected total reward of the policy, such as PPO (proximal policy optimization)\cite{2017ppo} and DDPG (deep deterministic policy gradient)\cite{2015ddpg}.  
They are used in re-ranking methods for generator training, such as \cite{2023GRN}, \cite{2023CMR}, \cite{2025NLGR}.
However, reward bias may cause algorithms to fail to converge or converge to a local optimum\cite{2015High}.
In recent LLM alignment, the evolution from PPO and GRPO (group relative policy optimization)\cite{2024grpo} to direct preference optimization methods bypasses reward modeling via direct human preference alignment through contrastive objectives. 
DPO\cite{2024dpo, 2025Gref} optimizes generative re-ranking by refining sequence probability distributions based on preference comparisons without specific rewards.
Some other methods incorporate Supervised Fine-tuning with preference optimization, such as ORPO\cite{2024orpo}, SimPO\cite{2024simpo}, and CHORD\cite{2025CHORD}.

\section{Problem Definition}
In industrial recommendation systems, a user's continuous browsing is grouped into one session. 
A session comprises multiple requests, each triggering the full recommendation pipeline.
% : recall, pre-ranking, ranking, and re-ranking.
When a request $q$ comes, the user’s recent interaction history and profile features, e.g., user ID, gender, are used for recommendation.
% the current state.
% If some requests in a session are delayed due to interest updates or session timeouts, suboptimal recommendations can result
Re-ranking adjusts candidate orders ($\mathcal I = [i_1, i_2, ..., i_N]$) from upstream to construct contextually optimal permutations ($\tau^* = [g_1, g_2, ..., g_M], M \leq N$) consisting of M items to be exposed to the user.
Given upstream limitation due to cold start, interest delay, data drift, and timeout problems, our goal is first to train a reward model $R_{\theta} (\tau, q)$ consisting of immediate and exploratory rewards
and then solve the optimal sequence $\tau ^* $ to maximize $R_{\theta} (\tau, q)$.
\begin{equation}
    \tau ^* = \underset{\tau \in \Omega}{\arg\max} \, \mathbb{E}\left[R_{\theta}(\tau, q)\right]
\end{equation}
, where $\Omega \ ( |\Omega| = A^M_N)$ is all permutation space from upstream and $\theta \in \Theta_R$ is learnable parameters of reward model.
During online serving, our generator directly generates the final sequence, instead of first generating and then evaluating in the GE framework\cite{2023PIER,2025NAR4Rec}.

\section{METHODOLOGY}
In this section, we present a \textbf{Dual Exploratory-Driven Generative Re-ranking Model} (DEGR), a framework that leverages a \textbf{hybrid supervised-reinforcement exploration and optimization paradigm} guided by the \textbf{exploratory reward model}, as illustrated in Figure\ref{fig:arch}.
Our reward model dynamically balances immediate and exploratory rewards, thereby preserving browsing potential and enabling serendipitous conversions.
Subsequently, it provides explicit signals to steer the generator's exploration.
The generator employs an encoder-decoder architecture, using multiple decoding-head cohorts to generate sequences efficiently.
The exploration and optimization paradigm integrates three components: supervised learning, exploration diversity constraint, and adaptive reward-weighted ORPO, which enables the generator to produce optimal sequences with higher exploratory reward, further serving as an adaptive cross-request contextual bridge.

\subsection{Exploratory Reward Model}
In our paradigm, the reward model holds a position of significance comparable to the evaluator in the Generator-Evaluator (GE) framework \cite{2025rerank_evaluator}. 
While the latter is responsible for final sequence selection, our reward model synthesizes efficiency, business, and diversity objectives and steers the generator's exploration trajectories.
Existing methods\cite{2025rerank_evaluator, 2023PIER} focus on maximizing immediate utility, e.g., CTR(click-through rate) or CVR(conversion rate), yet remain suboptimal under rigid upstream supply constraints.
Specifically, when all candidates exhibit low immediate value, existing methods fail to generate sequences with latent exploratory worth, leading to a local optimum, as illustrated in Seq2 of Figure\ref{fig:brief_next}.
Therefore, we design an exploratory reward model that uses pCTR predicted by upstream ranking as dynamic weights to adaptively balance \textbf{item-wise immediate} and \textbf{sequence-level exploratory value}.
It preserves browsing potential for future requests and enables serendipitous conversions during exploratory interactions, enhancing global gains, as demonstrated by Seq1 in Figure\ref{fig:brief_next}.
Subsequently, the exploratory reward model steers the
exploration of the generator, effectively shaping it into an adaptive cross-request contextual bridge under supply constraints.

\subsubsection{\textbf{Model Architecture}} 
As shown in Figure\ref{fig:arch}(B), the reward model extracts item embeddings through DIN\cite{2017DIN} to capture user behavior patterns.
After element-wise summation of item and sinusoidal position embeddings, it performs self-attention across all items in sequences.
During the reward model's training and generator's training, it captures authentic contextual awareness of exposure logs and generated sequences, respectively.
% in the sequence to model authentic inter-context relationships.
Then the context-aware embeddings drive two prediction tasks: item-wise immediate value ($R_{g1}, R_{g2}, ..., R_{gM}$) and sequence-level exploratory value ($R_{E}$) via multi-gate mixture-of-experts (MMoE) \cite{2018mmoe} and multilayer perceptron (MLP), respectively.

\begin{figure}
  \centering
  \includegraphics[width=0.46\textwidth, height=2.9cm]{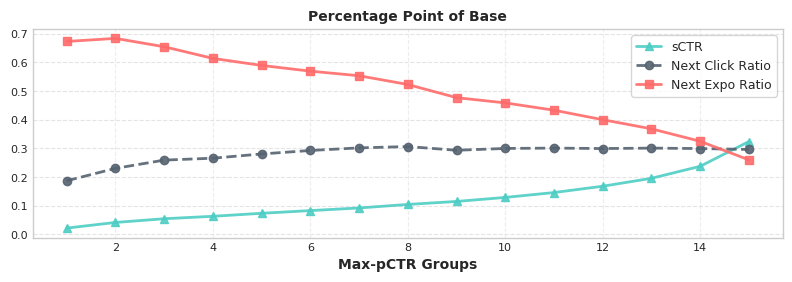}
  \caption{
  Relationships between Max-pCTR and sCTR, Next Click, and Next Expo Ratio in JD recommendation scenarios.
}
\label{fig:ana}
\end{figure}

\subsubsection{\textbf{Optimization Objective}}
Existing re-ranking primarily optimizes for immediate consumption utility, e.g., clicks or purchases, which proves insufficient under constrained upstream supply, regardless of sequence permutations.
This limitation is empirically observed in our JD recommendation scenario in Figure\ref{fig:ana}.
We define \textit{Max-pCTR} as the maximum predicted CTR (predicted by upstream ranking) within the sequence, 
\textit{sCTR} as the sequence-level click-through rate (the probability of at least one item receiving a click in sequence), \textit{Next Click Ratio (NCR)} as the probability of producing a click after the current request, and \textit{Next Expo Ratio (NER)} as the probability of continuing to browse after the current request.
The subsequent improvement in NCR and NER demonstrates that our re-ranking can serve as an adaptive cross-request bridge.
Note that Max-pCTR can get in the current request, while sCTR, NCR, and NER are posterior statistics.
sCTR will be affected by the quality of upstream supply and item orders in sequence, while Max-pCTR more purely reflects the quality of upstream supply, where a lower Max-pCTR reflects a low-quality supply.
Our analysis, conducted across 15 equal-frequency groups of Max-pCTR, reveals a statistically significant positive correlation between Max-pCTR and sCTR.
Crucially, we find that Max-pCTR and NCR are essentially unrelated. 
This decoupling suggests that when immediate gains are limited by low-quality supply, we can strategically prioritize \textit{NER} to encourage deeper exploration.
That is the reason why re-ranking should adaptively balance immediate gains and personalized exploration, further preserving exposure opportunities to cultivate latent conversion potential.
Consequently, we propose an exploratory reward model that leverages pCTR predicted by the upstream to balance item-wise immediate and sequence-level exploratory value.

\textbf{1. Item-wise Immediate Reward.}
% We predict item-wise immediate value, such as click-through rate (CTR) or conversion rate (CVR) for each item based on the binary cross-entropy (BCE) loss.
We predict immediate value for each item, e.g., CTR, CVR, by binary cross-entropy (BCE) loss.
% like traditional methods  
\begin{equation}
    L_{p} = \sum_{ j } - y_j \cdot log(R_j) - (1-y_j) \cdot  log( 1-R_j), \ j \in \{g_1, g_2, ..., g_M\} 
\end{equation}
, where $y_j$ is the ground truth of whether to click or purchase, and $R_j$ is the predicted item-wise immediate score of item $j$.

\textbf{2. Sequence-level Exploratory Reward.}
The sequence-level exploratory reward $R_{E}$ is derived by aggregating embeddings followed by an MLP to capture a global sequence characteristic.
We cast the exploratory reward prediction as a binary classification task as Eq.\ref{eq:ls}, leveraging a five-tier sampling strategy (labeled A through E). 
These categories are derived from a composite of current clicks and scrolls, and subsequent click actions, as detailed in Table~\ref{tab:sample}.
\begin{equation}
    L_{s} = - w_{\{*\}} \cdot y_e \cdot log(R_e) - w_{\{*\}} \cdot (1-y_e) \cdot  log( 1-R_e)
\label{eq:ls} 
\end{equation}
, where $w_{\{*\}}$ are different label weights of samples A $\sim$ E.
Samples D and E demonstrate user interest in the current exposure sequence, while sample E further reveals a sustained exploration motive; we set $w_E > w_D$.
Despite a lack of explicit interest, samples B and C serve as bridges, guiding the user toward further exploration, and C triggers engagement in subsequent exploratory interactions.
To encourage further exploration under low Max-pCTR of upstream constraints, we implement adaptive exploratory exposure boosting for future requests to cultivate latent conversions by dynamically adjusting training weights of sample C according to Eq.\ref {eq:weight}.
\begin{equation}
    w_C = ( \ 4 - 3 \log_2(1+\max_{1 \leq i \leq N} pCTR_j)\  ) \cdot w_C', \ pCTR_j \in [0,1]
\label{eq:weight}
\end{equation}
It increases positive weights as Max-pCTR decreases, prioritizing exploratory exposure.
% to cultivate latent conversion potential through sustained browsing depth.
Since category B samples have exploratory value under supply-constrained conditions, we mask sample B when Max-pCTR $\leq$ 0.01 during training to eliminate negative penalties, thereby promoting scroll and follow-up action implicitly.
The differentiation and dynamic weighting of the above samples achieve exploratory value modeling.
The final composite reward $R_\theta(\tau, q)$ of sequence $\tau = [g_1,...g_M]$ for request $q$ combines immediate $R_j$ and exploratory $R_{e}$ value through a weighted summation in Eq.\ref{eq:reward}.
\begin{equation}
    R_\theta(\tau, q) = \alpha \cdot R_{e}  + \sum_{j \in \{g_1,...g_M\} }  (1-\alpha ) \cdot \delta_j \cdot R_j  \ , \ \delta_j = 1-\log_{M+1}(j)
\label{eq:reward}
\end{equation}
$\delta_j$ represents the item position weight, which tends to decrease monotonically in practice. 
Parameters of the reward model are fixed to guide generator training, as described in Section 4.2.

\begin{table}
  \caption{Sequence-level label divisions based on future action}
  \label{tab:sample}
  \begin{tabular} {p{0.12cm}|p{2.17cm}|p{0.61cm}|p{0.65cm}|p{1.94cm}|p{1.1cm}}
    \toprule
    % Methods &  NDCG@K & MAP@K &$\tau$ & Click-AUC & Order-AUC\\
      & Description & Click & Scroll & Follow-up click & Label $y_e$\\
    \midrule
    A   &Terminate   & \ding{55} & \ding{55} & - & 0 \\
    B   &Invalid Explore  & \ding{55} & \ding{51} & \ding{55}  & 0 \\
    C   &Potential Interest & \ding{55} & \ding{51} & \ding{51} & 1 \\
    D   &Instant Match  & \ding{51} & \ding{55} & - & 1 \\
    E   &Deep Explore  & \ding{51} & \ding{51} & -  & 1  \\
    \bottomrule
  \end{tabular}
   \begin{tablenotes}
     \item[1] Category B is treated as negatives when Max-pCTR>0.01, otherwise masked to prevent noise.
     % Sequence click occurs when any item in sequence is clicked, and scrolling indicates the last item is exposed.
   \end{tablenotes}
\end{table}
\subsection{Dual Exploratory-Driven Generator}
Our generator leverages a hybrid supervised-reinforcement exploration and optimization paradigm guided by the exploratory reward model.
In this generator-only architecture, the reward model provides explicit signals to steer reinforcement learning (RL) exploration.
Based on this optimization paradigm, the generator learns to produce optimal sequences with higher exploratory reward, ultimately serving as an adaptive contextual bridge across requests.
% The generator acquires authentic contextual awareness when generating sequences.

Specifically, we employ an efficient encoder-decoder architecture, where the encoder extracts candidate representations, and the decoder utilizes multiple decoding-head cohorts to facilitate the efficient parallel generation of items.
% To optimize the generator, we propose a hybrid supervised-reinforcement learning exploration and optimization paradigm comprising three objectives:
The proposed hybrid supervised-reinforcement exploration and optimization paradigm comprises three objectives:
1. \textbf{Supervised Learning (SL)} aligns the generator with the online distribution and avoids potential capability collapse in RL.
2. \textbf{Exploration Diversity Constraint} achieves intra-cohort regularization when item parallel generation, effectively mitigating semantic redundancy and representation collapse across parallel decoding cohorts.
3. \textbf{Adaptive Reward-weighted ORPO} (AR-ORPO) guides the generator in exploring the sequence space to maximize exploratory rewards based on ORPO\cite{2024orpo}. 
It constructs preference lists and dynamically uses rewards as credible soft weights to refine the optimization trajectory.
% maximizes reward utility while suppressing noise.
Furthermore, to enhance exploration capacity in RL, we introduce a \textbf{Multi-Mechanism Sampling Strategy} to yield a diverse set of candidate sequences.

\subsubsection{\textbf{Efficient Encoder-Decoder}}
Our generator employs an end-to-end efficient encoder-decoder architecture, as illustrated in Figure\ref{fig:arch}. 
G-Encoder extracts candidate context representations across items and user behaviors through attention mechanisms.
% G-Decoder exclusively focuses on auto-regressive sequence generation via similarity-based incremental decoding.
% Additionally, we utilize decoding-head cohorts to achieve efficient decoding.
G-Decoder adopts a representation-aware incremental decoding mechanism for sequence generation, leveraging multiple decoding-head cohorts to achieve efficient parallel inference.

\textbf{G-Encoder.}
The G-encoder employs DIN\cite{2017DIN} to derive user-interest representations and utilizes the self-attention mechanism\cite{2017attention} to model global contextual interdependencies between item candidates, enabling comprehensive feature extraction and relationship modeling.
% Additionally, positional encoding is used for context-aware modeling of upstream sequence order.
% Its inclusion considers upstream sequence order, where upstream stages sometimes will take into account diversity, while its omission makes re-ranking focus exclusively on semantic contextual modeling of item attributes. 
It outputs contextually enriched embeddings for each candidate item and feeds them into the G-Decoder.

\textbf{G-Decoder.}
% As shown in Figure\ref{fig:arch}, based on multi-head attention mechanism, 
G-Decoder regards the G-Encoder's output embeddings as keys and values and previously generated embeddings as queries in a stepwise sequence expansion, thereby gradually synthesizing new embeddings ($\mathbf{emb'_{1}} \rightarrow \mathbf{emb'_{M}}$) until $M$ embeddings have been generated.
It initiates auto-regressive generation with a pre-defined fixed <start> embedding as the initial query. 
We also incorporate sinusoidal position embeddings used in the reward model into the query embeddings to make the generator acquire authentic contextual awareness when generating sequences.
% Through multi-head attention mechanisms, it regards the G-Encoder's output embeddings as keys and values and the <start> or generated embeddings as the query in a stepwise sequence expansion, thereby gradually synthesizing new embeddings ($emb_{p1} \rightarrow emb_{pM}$) until the generated embedding reaches $M$.
To accelerate decoding, we utilize decoding-head cohorts for parallel inference of looking ahead K items at a time.
After generating embeddings, G-Decoder computes dot-product similarities between them and G-Encoder's output embeddings and generates a similarity matrix for sequence sampling.  
% an $M\times N$ 
Row-wise softmax normalization yields probability distributions $P$, where element $(i,j)$ represents the likelihood of the i-th position corresponding to candidate $item_j$.

\textbullet{} \textbf{{Multiple Decoding-head Cohorts.}}
We use decoding-head cohorts with multiple feed-forward networks (FFN) and residual connection modules to accelerate decoding and parallel generation.
Each head predicts K continuation candidates for different future positions per decoding step, inspired by MTP\cite{2024mtp} and MEDUSA\cite{2024MEDUSA}.
\begin{equation}
\mathbf{emb'_{{i+k}}} = W^{(k)}_2 \cdot ReLU(W^{(k)}_1 \cdot h'_{{i}}) + h'_{{i}} \ , \ k \in \{1,...,K\}
\end{equation}
, where $h'_{i}$ is the current hidden state calculated from $\mathbf{emb'_{{i}}}$ at position $i$ based on the attention mechanism, and $\mathbf{emb'_{{i+k}}}$ is the output embedding of the decoding head $k$.
It reduces generation complexity from $O(M)$ to $O(M/K)$, achieving up to $K×$ speedup and enabling real-time industrial deployment.
% While inter-group context can be well modeled, intra-group context-aware information is partially degraded. 
% The cohort size $K$ is determined via ablation studies to establish a Pareto-optimal balance between efficiency and speed.
We observe that in re-ranking, where the candidate diversity is inherently limited, the multiple decoding-head cohorts tend to produce highly correlated embeddings. 
Thus, we introduce intra-cohort regularization as a structural constraint to mitigate semantic redundancy and ensure the distinctness of parallelly generated items in Section 4.2.2.

\subsubsection{\textbf{Hybrid Exploration and Optimization paradigm}}
% 环境交互解决ORPO探索局限
We propose a hybrid supervised-reinforcement learning optimization and exploration paradigm to explore and discover optimal, diverse sequences within the policy space that yield high exploratory rewards.
By coupling supervised stability with RL, this dual exploration enables the generator to function as an adaptive contextual bridge across requests.
It integrates the following three objectives.
Additionally, to enhance exploration capacity in RL, we propose multi-mechanism sampling to yield diverse candidate sequences.

\textbullet{} \textbf{{Supervised Learning.}} It fits online distribution and avoids capability collapse in RL based on categorical cross-entropy loss.
\begin{equation}
    L_{CE} = \sum_{i=1}^M \sum_{j=1}^N y_{ij} \cdot \log p_{ij} , \ p_{ij} = P(y_i=j \mid y_{<i}, \ q) 
\end{equation}
, where $q$ is the current request, $y_{ij} \in \{0,1\}$ indicates a match between i-th position and candidate $item_j$ of the online sequence as the ground-truth, and $p_{ij}$ is its predicted probability.

\textbullet{} \textbf{{Exploration Diversity Constraint (EDC).}} 
In re-ranking, where candidate diversity is inherently limited, we observe that multiple cohorts tend to produce highly correlated items. 
We introduce \textbf{intra-cohort regularization} as exploration diversity constraints to effectively mitigate semantic redundancy and representation collapse, ensuring the distinctness of parallelly generated items.
\begin{equation}
    L_{ICR} = \sum_{j=1}^{K-1} \left( \frac{1}{K-j}\sum_{k=j+1}^K\left( \frac{\mathbf{emb'_{{i+j}}}^\top \mathbf{emb'_{{i+k}}}}{\|\mathbf{emb'_{{i+j}}}\|_2 \|\mathbf{emb'_{{i+k}}}\|_2} \right)^2 \right)
\end{equation}
, which calculates the similarity of generated embeddings in cohorts.

\textbullet{} \textbf{{Adaptive Reward-weighted ORPO (AR-ORPO).}}
Supervised learning fits online distribution that may not quite meet multiple objectives, while reinforcement learning can guide models in exploring the sequence space to maximize our exploratory rewards. 
We introduce Adaptive Reward-weighted ORPO (AR-ORPO) loss based on ORPO\cite{2024orpo}.
It constructs \textbf{preference lists} over mere pairs and dynamically uses rewards as \textbf{credible soft weights} to refine optimization, maximizing reward utilities of sampled trajectories while suppressing noise.
For each sequence $\tau$ in the sampled set $\mathcal T=\{\tau_1, \tau_2, ..., \tau_N\}$ of request $q$, we calculate sequence probability $P(\tau \mid q)$ and exploratory reward $R_{\theta}(\tau, \ q)$ based on reward model.
\begin{algorithm}
\caption{Procedure for Group Beam Search.}
\label{alg:group_beam_search}
\begin{algorithmic}[1]
\State Input: Generated probability $P(y_i \mid y_{<i}, q)$ with Gumbel noise, beam size $b$, group size $g$, to be generated sequence length $M$.
\State $ \mathcal T_g \gets \emptyset$ \Comment Initialize sampled trajectories.
% \State $p_1 \gets P(y_1 \mid q)$  \Comment get the probability at the first step.
% \State $G_g, P_g$ $\gets$ Top($p_1$, g) \Comment select $g$ subsequences $G_g$ with top probabilities $P_g$
\State $ \mathcal T_g, P_g \gets  Top \left( P(y_1 \mid q) , \ g \right)$ \Comment Sample $g$ start items $ \mathcal T_g = \{ \tau \}, \ (|\mathcal T_g| = g)$ with top probabilities $P_g$.

\State for $i = 2,3,..., M$ do:
\State \ \ \ \ $p_i \gets P(y_i \mid y_{<i}, q)$  \Comment Get current step's probability
\State \ \ \ \ for $k = 1,2,..., g$ do: \Comment For each group do beam search
\State \ \ \ \ \ \ \ \ $P^{(k)}_g \gets MASK(p_i,\ \mathcal T^{(k)}_g) \cdot P^{(k)}_g$  \Comment mask $p_{ij}$ as 0 at selected position $j \in \mathcal T^{(k)}_g[:,:i-1]$ and compute sequence probability.
\State \ \ \ \ \ \ \ \  $\mathcal T_g^{(k)}, P^{(k)}_g \gets Top(P^{(k)}_g, b) $ \Comment for each group, select $b$ subsequences $\mathcal T_g^{(k)}, \ (|\mathcal T_g^{(k)}| = b )$ with top probabilities
\State $\mathcal T_g \gets \bigcup^g  \mathcal T_g^{(k)}$, $P_g \gets \bigcup^g P_g^{(k)}$ \Comment merge sequences.

\State Return: $\mathcal T_g, P_g$ \Comment Return sampled sequences and probabilities.
\end{algorithmic}
\end{algorithm}
\begin{equation}
P(\tau \mid q) = \prod_{i=1}^{M}  P(y_i=g_i \mid y_{<i}, q) , \ \tau = [g_1, g_2, ..., g_M]
\end{equation}
We order sampled trajectories $\mathcal T$ by the reward $R_{\theta}(\tau, \ q)$ in descending order, 
and then sample S quantiles to obtain more representative sequences $[\tau_1, \tau_2, ..., \tau_K]$ and rewards ($R_{\theta}(\tau_1, \ q) > R_{\theta}(\tau_2, \ q) > ... > R_{\theta}(\tau_K, \ q) $).
The most favored and disfavored sequences is $\tau_1$ and $\tau_K$, respectively.
The AR-ORPO loss is defined as:
\begin{equation}
    L_{A} = \sum_{i=1}^{S-1} \underbrace{\frac{e^{R_{\theta}(\tau_i, \ q) / t }}{ \sum_{j=1}^{K} e^{R_{\theta}(\tau_j, \ q) / t}  }}_{soft \ \ weights} \cdot \log \sigma \left( \log \frac{odds(\tau_i \mid q)}{\sum_{j=i+1}^{S} odds(\tau_j \mid q)} \right)
\end{equation}
, where $odds(\tau \mid q)=P(\tau \mid q)/(1-p(\tau \mid q))$.
We compute soft weights by applying the softmax function to sequence rewards, leveraging the importance sampling strategy.
It ensures the favored sequence is assigned an odds ratio sufficiently high to exceed the cumulative odds of lower-ranked sequences.
$t$ is a temperature parameter that controls weight distribution. 
A smaller $t \to 0 $ makes the weight approach 1.
% \begin{equation}
%     L_{A} = \underbrace{\frac{e^{R_{\theta}(\tau_w, \ q) / t }}{ e^{R_{\theta}(\tau_w, \ q) / t} + e^{R_{\theta}(\tau_l, \ q) / t } }}_{soft \ \ weights} \cdot \log \sigma \left( \log \frac{odds(\tau_w \mid q)}{odds(\tau_l \mid q)} \right)
% \end{equation}
% , while a disfavored sequence is opposite.
The final loss of the generator is as Eq.\ref{eq:l_gen}.

\begin{equation}
L_{gen} = \frac{1}{|Q|} \sum_{q \in Q} \left( L_{CE} + \beta \cdot L_{ICR} + \gamma \cdot L_{A} \right)
\label{eq:l_gen}
\end{equation}
After training, DEGR produces sequences with high exploratory rewards, becoming an adaptive cross-request contextual bridge.

% \subsubsection{\textbf{Multi-Mechanism Probabilistic Sampling.}}
\textbullet{} \textbf{{Multi-Mechanism Probabilistic Sampling.}}
To improve exploration capacity in RL, we introduce multi-mechanism sampling that integrates group beam search and heuristic sampling methods to generate diverse trajectories in the probability space and prevents homogenization.
Diverse trajectory-reward mappings and broader reward landscapes enhance exploration efficiency and prevent models from being trapped in local optima, thereby facilitating preference optimization.
% This exploration offers diverse sequence-reward mappings and broader reward landscapes, thereby facilitating preference optimization, refining the model's generative ability, and preventing the model from becoming trapped in local optima. 
During online serving, we generate one sequence with the highest probability based on greedy sampling.
% , since the trained generator learns preferences from the reward model.
% directly use greedy sampling without noise

\textbf{Group Beam Search.}
For each step $i$, we calculate probabilities $P(y_i\mid y_{<i}, q)$ with Gumbel noise based on generated $logits(y_i|y_{<i}, q)$.
% , converging to homogenized outputs due to greedy policy.
\begin{align}
    &g = - \log( - \log(u)),\ u \sim Uniform(0,1)  \\
    &P(y_i\mid y_{<i}, q) = softmax(logits(y_i|y_{<i}, p) + g) ,\ i \leq M \nonumber 
% \label{eq:gumbel}
\end{align}
\begin{algorithm}
\caption{Procedure for Heuristic Sampling.}
\label{alg:heuristic}
\begin{algorithmic}[1]
\State Input: Generated probability $P(y_i \mid y_{<i}, q)$ with Gumbel noise, Sampled sequences $\mathcal T_h$ by heuristic sampling.

\State $P_h \gets 1$
\State for $i = 1,2,..., M$ do:
% \State \ \ \ \ $p_t \gets $  \Comment get current step's probability.
\State \ \ \ \ $p_t \gets gather\left(P(y_i|y_{<i}, q), \mathcal T_h[:,i]\right)$ \Comment get current step's probability of heuristic sampled sequences.
\State \ \ \ \ $P_h \gets MASK(p_i, \mathcal T_h) \cdot P_h$  \Comment mask $p_{ij}$ as 0 at selected position $j \in \mathcal T_h[:,:i-1]$ and compute sequence probability $P_h$.
% \Comment put $0$ at selected positions of $p_i$ for masking and then compute probabilities of sequences.

\State Return: $P_h$ \Comment probabilities of $\mathcal T_h$.
\end{algorithmic}
\end{algorithm}
Beam search may suffer from a diversity deficiency and converge to homogenized outputs\cite{2016Diverse_Beam_Search}.
To improve diversity, we utilize group beam search to maintain different starting items in sampled trajectories, as described in Algorithm\ref{alg:group_beam_search}. 
To ensure sequences do not contain identical items, we mask selected items during sampling by setting the probability of the corresponding position to $0$.

\textbf{Heuristic Sampling.}
Building upon heuristic sampling paradigm of traditional GE architecture, we first sample sequences by dynamically weighting item scores predicted by ranking models (e.g., CTR, CVR) to fuse different preferences and injecting stochastic perturbations.
We incorporate the exposure sequence with the above sampled sequences to obtain $\mathcal T_h$, and calculate corresponding probabilities based on Algorithm\ref {alg:heuristic}.
The final sequence set through our multi-mechanism sampling method is denoted as $\mathcal T = \mathcal T_g \cup \mathcal T_h$.

\begin{table*}
  \caption{Comparison of Overall Performance on the Public Taobao Dataset and JD Production Dataset}
  \label{tab:all_result}
  \begin{tabular}{p{1.5cm}|p{1.2cm}p{1.2cm}p{1.2cm}p{1.2cm}|p{1.2cm}p{1.2cm}p{1.2cm}p{1.2cm}p{1.2cm}p{1.2cm}}
  % \begin{tabular}{p{0.25cm}p{1.85cm}|cccc|cccc|cccc|cccc}
    \toprule
    % Methods &  NDCG@K & MAP@K &$\tau$ & Click-AUC & Order-AUC\\
    {\multirow{2}{*}{Method}} & \multicolumn{4}{c|}{Taobao}&\multicolumn{6}{c}{JD}   \\
    % \cline{3-8}
    {} &GAUC & NDCG & MAP@2 & Recall@2 & GAUC  & NDCG & MAP@2 & MAP@4  & Recall@2 & Recall@4 \\
         
    \midrule 
    DCN & 0.5870 & 0.1107 & 0.0835  & 0.0975 & 0.6255 & 0.7376 & 0.5748 & 0.6341 & 0.6673 & 0.8726\\
    PRM & 0.5983 & 0.1209 & 0.0847  & 0.0989 & 0.6380 & 0.7436 & 0.5850 & 0.6427 & 0.6756 & 0.8775\\
    \midrule
    PIER & 0.6004 & 0.1221 &0.0851 & 0.1005 &0.6387 & 0.7442 & 0.5863 & 0.6430 & 0.6765  & 0.8793 \\
    GRN & 0.6015 & 0.1232 & 0.0859  & 0.1011 & 0.6385 & 0.7446 & 0.5879 & 0.6437 & 0.6766 & 0.8808 \\
    CMR & 0.6025 & 0.1246 & 0.0863  & 0.1023 & 0.6393 & 0.7449 & 0.5883 & 0.6441 & 0.6772 & 0.8812 \\
    NAR4Rec & 0.6016 & 0.1233 & 0.0858  & 0.1012  & 0.6364 & 0.7442 & 0.5875 & 0.6432 & 0.6770 & 0.8800 \\
    MG-E(G=4) & 0.6031 &0.1249 &0.0865 &0.1029 & 0.6392 & 0.7454 & 0.5891 & 0.6448 & 0.6784 & 0.8822 \\
    GReF & 0.6037 & 0.1248 & 0.0866 & 0.1036 & 0.6403 & 0.7457 & 0.5901 & 0.6452 & 0.6791  & 0.8826\\
    DEGR & \underline{0.6107} & \underline{0.1287} & \underline{0.0871}  & \underline{0.1082}  &\underline{0.6486} &\underline{0.7493} &\underline{0.5951} &\underline{0.6505} &\underline{0.6839} &\underline{0.8871}   \\ 
    \bottomrule
  \end{tabular}
\end{table*}

\section{EXPERIMENT}
We evaluate our method through a series of offline experiments and online A/B testing to answer the following questions.
\begin{itemize}[leftmargin=*]
\item \textbf{Q1}: Does DEGR enhance the overall performance by adaptively balancing exploratory reward with immediate gains?
\item \textbf{Q2}: Does DEGR dynamically construct a cross-request contextual bridge by the dual exploration?
\item \textbf{Q3}: How does each module in DEGR perform in performance?
\item \textbf{Q4}: Does the complexity of DEGR adapt to industrial systems?
\end{itemize}

\subsection{Experiment Setup}
\subsubsection{Dataset} 
% The used dataset includes the following two datasets:
% \textbf{Taobao Public Dataset}: It is a public dataset\footnote{https://tianchi.aliyun.com/dataset/56} with 26 million expression
% logs of 1 million users and 0.8 million items in 8 days.
% The data from the first 7 days and the last 1 day are used as training and validation datasets, respectively.
\textbf{Public Taobao Dataset} contains 26 million expression logs in 8 days, split 7:1 for training–test. 
% We group items by user and timestamp as a single request.
We group up to 10 items per user-timestamp combination as a request.
\textbf{JD Production Dataset} contains one billion requests of one hundred million users in JD homepage recommendation, split 8:1 for training–test. 
% We take the homepage requests two days later as the validation set.

\subsubsection{Metrics}
% The offline and online experiments utilize the following metrics to evaluate the performance: 
% We utilize the following metrics to evaluate the performance.
\textbf{Offline evaluation metrics}: 
% Offline performance is benchmarked across the following metrics:
% Offline performance is evaluated on the following metrics:
\begin{itemize}[leftmargin=*]
\item GAUC (Area Under ROC Curve weighted by \#clicks)\cite{2017DIN} is a metric that measures the personalized discriminative ability for individual users' positive and negative samples.
In essence, it can be computed from relative ranking results output by our generator without the need for specific prediction scores.
\item NDCG (Normalized Discounted Cumulative Gain)\cite{2019Seq2Slate,2023GRN,2025NLGR,2025NAR4Rec}: to assess position-aware ranking quality for top-ranked items.
\item MAP@K (Mean Average Precision)\cite{2019Seq2Slate, 2023CMR}: 
% ranking precision.
to emphasize overall ranking quality of all relevant terms in the topK terms.
% \item HR@K\cite{2019ExactK} and 
\item Recall@K\cite{2025NAR4Rec}: to quantify user engagement coverage on topK recommendations, regardless of exact order.
% , as formulated below.
\end{itemize}
% $$
% Recall@K = \frac{1}{|U|} \sum_{u \in U} \frac{|A_k \cap I_u|}{|I_u|}
% % HR@K = \frac{1}{|U|} \sum_{u \in U} \frac{|A_k \cap I_u|}{K} ; \ 
% $$
% , where $I_u$ denotes the click or purchase sets and $A_K$ is the topK item set of the generated sequence.
In practice, for Taobao, $K=2$ is used for MAP and Recall due to fewer interaction items in some requests, while JD supports both $K=2$ and $K=4$.
Higher GAUC, NDCG, MAP, and Recall indicate better performance in offline evaluation.
However, due to exposure bias in historical data, new recommended items may lack real feedback, so offline metrics do not guarantee online alignment.
Additionally, for reward model evaluation, traditional metrics like AUC evaluate overall sequence quality, which do not directly translate to online metrics.
To address this, we generate $16$ sequences via heuristic sampling based on ranking scores, then select the optimal one via reward models. 
It enables \textit{fair comparison between reward models and generators under consistent criteria}.
However, due to limited candidate coverage from heuristics, the reward model's performance typically lags slightly behind that of the generator.

\textbf{Online metrics}:
We use UCTR (click PV/exposure UV) and PV (exposure PV)  to evaluate online performance in JD. 
A higher PV indicates deeper user exploration and broader item exposure.

\subsubsection{Baselines} 
We compare our dual exploration-driven generator with both one-stage (DCN\cite{2017DCN}, PRM\cite{2019prm}), two-stage (PIER\cite{2023PIER}, GRN\cite{2023GRN}, CMR\cite{2023CMR}, NAR4Rec\cite{2025NAR4Rec}) and MG-E\cite{2025MG_E} and generator-only (GReF\cite{2025Gref}) methods as baselines. 
DCN incorporates cross-features at each layer, and PRM models contextual information among items based on self-attention and greedily generates sequences.
PIER unifies the generation and evaluation modules into a single model and is trained in an end-to-end manner.
GRN and CMR train evaluators based on item-wise and list-wise values, and use a pointer network to generate sequences auto-regressively. 
NAR4Rec achieves a non-autoregressive sequence generation based on a matching model.
MG-E employs an explicit loss to increase discrepancies among generators to improve sequence diversity. Here, we set the number of generators as $4$.
GReF introduces OMTP to improve the generation efficiency and uses DPO to align users' real feedback without a reward model.
Beyond this, in ablation studies, we also design multiple versions of DEGR, including the reward model and the generator.

\subsubsection{Implementation Detail}
The objective of our re-ranking is to select an optimal sequence with $M=10$ items to be exposed to users from a set of $N=30$ item candidates.
% In the reward model and the encoder and decoder in the generator, 
In DEGR, we employ one attention layer in the transformer architecture. 
The item-wise score prediction in reward model involves click and purchase tasks, based on an MMoE model containing two experts with 128 units.
In final reward calculation, the weight $\alpha$ of sequence-level exploratory value is set as $0.2$.
To accelerate decoding in G-Decoder, we set $K=6$ FFN after the attention layer for parallel generation.
While each decoding step could theoretically produce $6$ items, we optimize for performance by executing three decoding passes that yield 1, 3, and 6 items. 
We limit the number of items generated in the first two steps, which can maintain high-fidelity contextual modeling for the most important (top) items.
In multi-mechanism sampling, we set $b=2, g=3$ for group beam search and sample $7$ sequences for heuristic sampling.
In the generator training loss, we set loss weights $\beta=2.0$ and $\gamma = 0.01$.
And in AR-ORPO loss, we set the temperature parameter $t=0.2$, decided by experiments.

\subsection{Offline Performance Comparison (Q1)}
% \subsubsection{Performance on Taobao Dataset}
% Since the public dataset lacks unexposed data, Table\ref{tab:res_taobao} can not fully reflect differences between Rethink and HCCP, we conduct experiments on JD production dataset.
% \subsubsection{Performance on JD Production Dataset}
As illustrated in Table\ref{tab:all_result}, our Dual Exploratory-Driven Generative Re-Ranking architecture (DEGR) outperforms all baseline methods in Taobao public and JD datasets, particularly on MAP@2 and MAP@4.
This demonstrates DEGR's superior capability for context-aware sequence generation. 
Compared to baselines, DEGR offers three key advantages.
(1) DEGR's reward model adaptively balances exploratory with immediate gains, enabling the construction of adaptive cross-request contextual bridges.
(2) Our hybrid exploration and optimization paradigm helps to guide the generator to produce optimal sequences.
To improve the \textbf{inter-sequence diversity}, MG-E employs an explicit loss to increase discrepancies among generators, while DEGR enhances diversity via multi-mechanism probabilistic sampling and maximizes the probability of the highest-reward sequence during training.
Compared with MG-E\cite{2025MG_E} and GReF\cite{2025Gref}, DEGR designs an exploratory reward and leverages intra-cohort regularization as an exploration diversity constraint to ensure \textbf{intra-sequence item diversity}, thereby demonstrating the efficacy of our proposed method.
(3) Instead of the non-autoregressive matching model\cite{2025NAR4Rec} or RNN-based pointer network\cite{2023GRN,2023CMR}, DEGR employs an encoder-decoder architecture that more effectively captures and utilizes contextual information.

\subsection{Online A/B Test (Q1 \& Q2)}
% 说明了点击随着曝光的增加而增加，
We conduct A/B tests on the JD homepage recommendation system for 7 days.
DEGR has been deployed online to serve the main user traffic, achieving a \textbf{1.22\%} increase in UCTR and a \textbf{0.20\%} increase in PV.
Additionally, we have also deployed \textit{DEGR(w/o EDC)} version (excluding the exploration diversity constraint) online, which yields only a 0.72\% increase in UCTR with 0.20\% in PV, which demonstrates the effectiveness of our exploration diversity constraint.
We find that the larger the upstream supply number (N), the more pronounced the effect of the exploration diversity constraint becomes.

\begin{figure}
  \centering
  \includegraphics[width=0.46\textwidth, height=2.6cm]{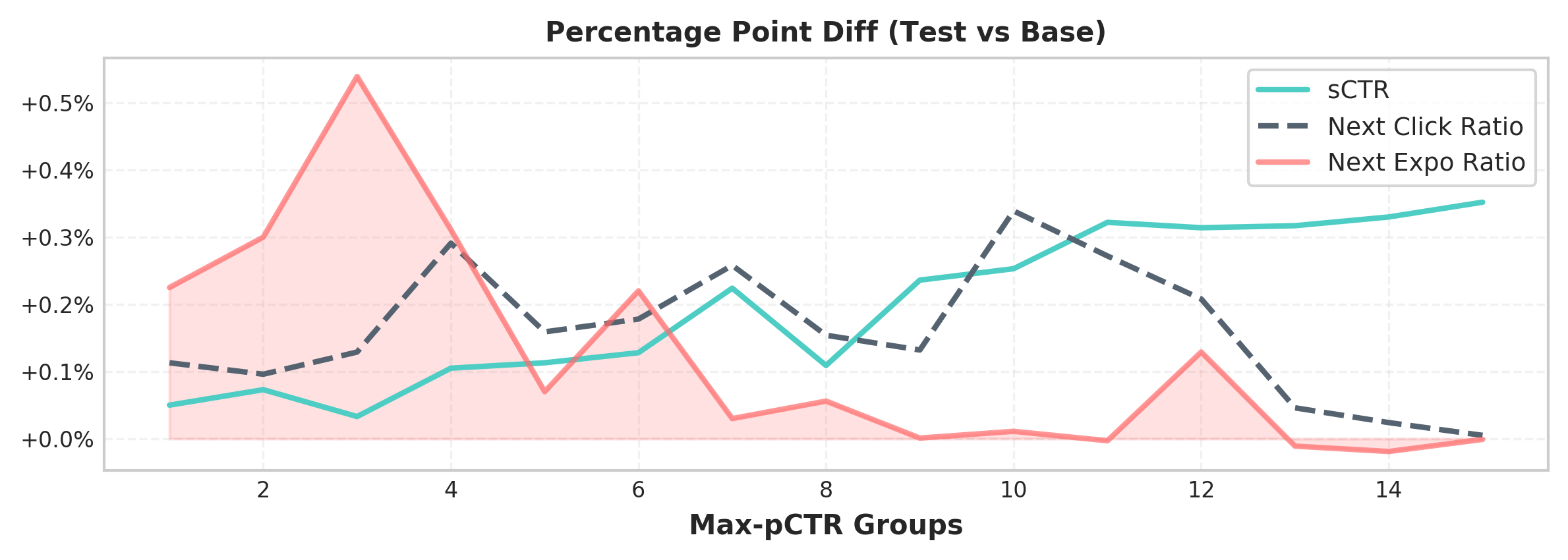}
  \caption{
  Relationships of Max-pCTR with relative improvements in sCTR, Next Click, and Next Expo Ratio between Test and Base. 
  The absolute value of Base is shown in Figure\ref{fig:ana}.
}
\label{fig:ana2}
\end{figure}

We also analyze the relationship between Max-pCTR and the relative improvement of sCTR, Next Click Ratio (NCR), and Next Expo Ratio (NER) between DEGR and our base PRM\cite{2019prm} model in Figure\ref{fig:ana2}.
A comprehensive evaluation shows that DEGR achieves significant improvements over the base in sCTR, NCR, and NER.
Under a low Max-pCTR caused by upstream constraints, DEGR enhances exploratory exposure (reflected in the improvement of NER) and further cultivates latent conversion potential (evidenced by the improvement in NCR).
Under a higher Max-pCTR when upstream models perform well, DEGR directly improves NCR with only mild gains in NER.
The improvement in NCR and NER demonstrates that DEGR achieves an adaptive balance between immediate and exploratory gains, and finally dynamically constructs an adaptive contextual bridge across requests.

\begin{table}
  \caption{Ablation Studies on JD Production Dataset}
  \label{tab:ablation}
  \begin{tabular} {p{2.23cm}|p{0.72cm}|p{0.72cm}|p{0.65cm}|p{0.65cm}|p{0.65cm}|p{0.65cm}}
    \toprule
    Method & GAUC & NDCG & M@2 & M@4 & R@2 & R@4 \\ 
    \midrule
    ER(w/o E)    &0.6388 &\underline{0.7334}  &0.5826 &0.6348 &0.6714 &0.8671   \\
    % ER(w/o DW)   &0.6386 &0.7333  &0.5831 &0.6349 &0.6714 &0.8675 \\
    ER           &\underline{0.6393} &\underline{0.7334}  &\underline{0.5834} &\underline{0.6351} &\underline{0.6718} &\underline{0.8681}  \\
    \hline
    DEGR(w/o E)  & 0.6447  &0.7475  &0.5926 &0.6481 &0.6818 & 0.8849  \\
    % DEGR(w/o DW) & -  &-  &- &-  &- & -  \\
    % \hline
    % DEGR(w/ BS)  & -  &-  &- &-  &- & -  \\
    DEGR(w/ GBS) &0.6453 &0.7472 &0.5913 &0.6476 &0.6803 &0.8863  \\
    % \hline
    DEGR(w/ ORPO) &0.6463 &0.7478 &0.5924 &0.6486 &0.6817 &0.8870  \\ 
    DEGR(w/o EDC) &0.6479 &0.7486 &0.5940 &0.6497 &0.6834 &\underline{0.8874}  \\ 
    % \hline  
    DEGR  &\underline{0.6486} &\underline{0.7493} &\underline{0.5951} &\underline{0.6505} &\underline{0.6839} &0.8871  \\
    \bottomrule 
  \end{tabular}
   \begin{tablenotes}
     \item[1] Here, M@K and R@K means MAP and Recall.
   \end{tablenotes}
\end{table}

\subsection{Ablation Studies (Q3)}
Ablation studies evaluate the effectiveness of each component of DEGR on JD production dataset, and results are displayed in Table\ref{tab:ablation}.

(1) \textbf{The Importance of Exploratory Reward}.
We compare the exploratory reward model (ER) and the generator(DEGR) guided by the reward model, respectively.
For the reward model, we directly compare ER with ER(w/o E) using the validation method described in Section 5.1.2.
The difference lies in whether they incorporate exploratory value.
Results indicate that the exploratory reward model can select effective sequences.
% Since sequences sampled by heuristic sampling are limited in coverage, regardless of which sequence the reward model selects, its performance slightly trails generators, especially in terms of top-ranked capability. 
For the generator, we compare DEGR (w/o E) and DEGR that are trained based on ER(w/o E) and ER.
Training guided by different reward models, DEGR outperforms DEGR(w/o E) by achieving a 0.25pp increase in Map@2 and 0.21pp in Recall@2, indicating the effects of exploratory rewards.

It is worth noting that although the reward model shows minimal improvement with and without the exploration reward, the final DEGR achieves notable improvement when the exploration reward is incorporated. 
This discrepancy is attributable to our evaluation protocol for reward models, which assesses sequence selection ability on sampled candidates produced by heuristic sampling, which are inherently limited in coverage and diversity.
In such a constrained setting, the effect of adding the exploration reward is largely muted, yielding only minor improvements in the reward model’s performance.
In contrast, the generator explores a much broader sequence space during training, allowing it to better exploit the exploration reward.

(2) \textbf{The Importance of Multi-Mechanism Sampling}.  
To evaluate multi-mechanism sampling, we compare DEGR with DEGR(w/ GBS) using group beam search (GBS) based on ER.
Multi-mechanism sampling provides a more diverse exploration space, which achieves an enhancement across all metrics compared to DEGR(w/ GBS).

(3) \textbf{The Importance of Exploration Diversity Constraint (EDC)}. 
A direct comparison between DEGR and DEGR(w/o EDC) reveals that the exploration diversity constraint yields a 0.11pp lift in MAP@2.
The online A/B test also demonstrates its efficacy.

\begin{figure}
  \centering
  \includegraphics[width=0.47\textwidth, height=2.8cm]{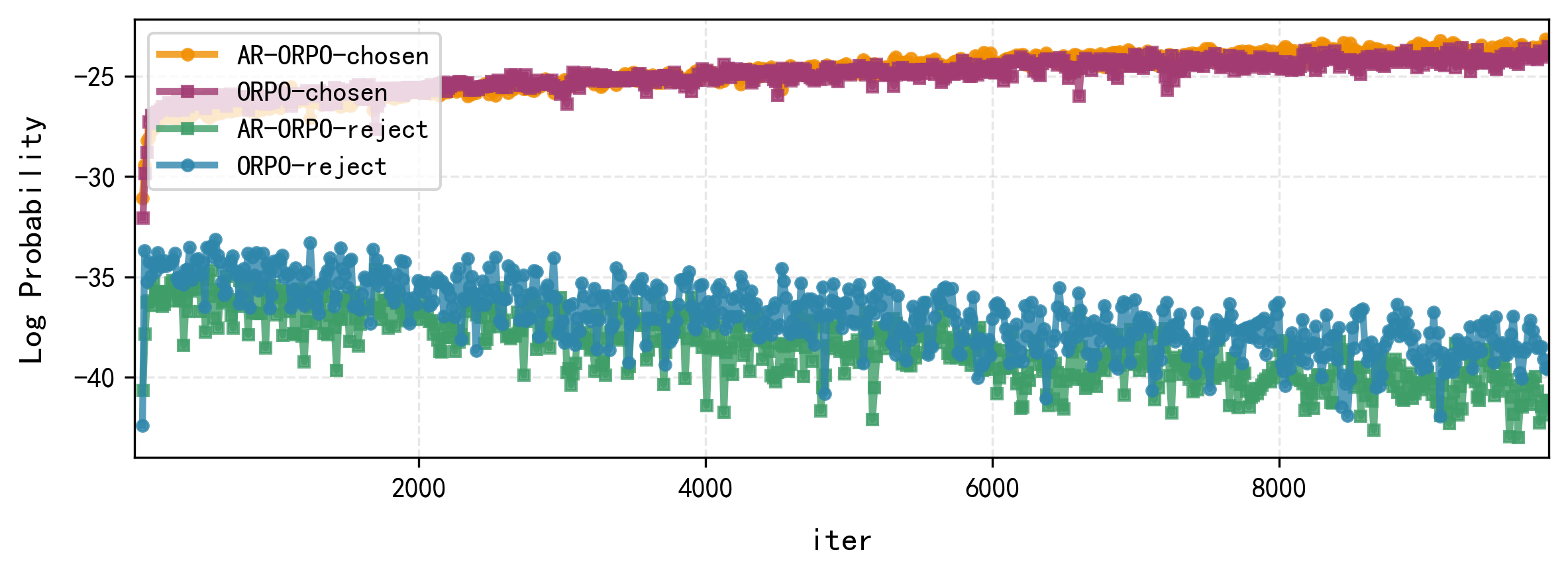}
  \caption{Log Probability of chosen and rejected sequences of AR-ORPO and ORPO.
}
\label{fig:prob}
\end{figure}

\begin{figure}
  \centering
  \includegraphics[width=0.46\textwidth, height=2.1cm]{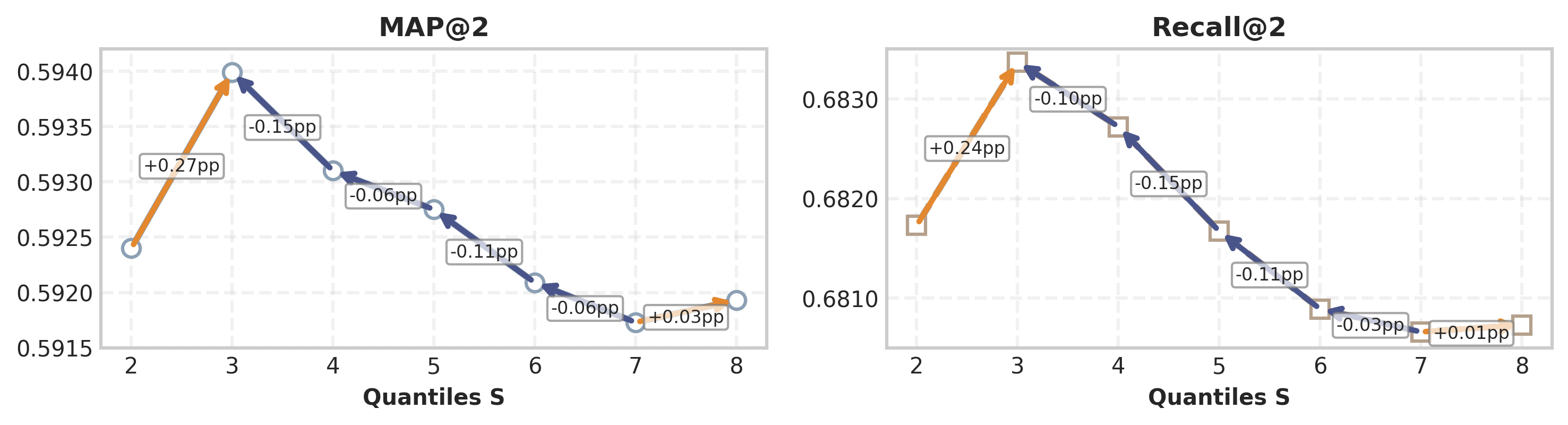}
  \caption{Quantile-Metric Relationships in AR-ORPO. Larger S may introduce noise, resulting in suboptimal performance.
}
\label{fig:orpo_trends}
\end{figure}

(4) \textbf{The Importance of Adaptive Reward-weighted ORPO (AR-ORPO)}. 
To evaluate the performance of our AR-ORPO loss, we compare DEGR(w/ ORPO) and DEGR(w/o EDC) optimized by ORPO and AR-ORPO loss, respectively. 
The result indicates that AR-ORPO loss achieves a 0.16pp rise in Map@2 and 0.17pp in Recall@2 by constructing preference lists over mere pairs and leveraging rewards as credible soft weights.
Contrasting probabilities and rewards of multiple divergent sequences maximizes reward utility and suppresses noise. 
% Additionally, our findings in Figure\ref{fig:orpo_trends} indicate that during training through dynamic ORPO, a greater number of quantiles $S$ does not invariably enhance outcomes.
% A higher $S$ value helps expand the explored sequence space, but it may introduce noise associated with the reward model's evaluations.
Figure\ref{fig:prob} shows that AR-ORPO can better distinguish between chosen and rejected samples compared to ORPO.
Figure\ref{fig:orpo_trends} shows that in AR-ORPO, a higher quantile $S$ helps expand the explored sequence space, but may introduce noise from reward evaluations, which may compromise results.
In our experimental setup, $S=3$ achieves the best.

\subsection{Complexity and Time Analysis(Q4)}
Table\ref{tab:complexity} shows that there's little parameter count difference between reward models (evaluators) in baselines and DEGR.
However, the parameter count of DEGR's generator is above that of others due to underlying architectures, such as transformer\cite{2017attention} and pointer network\cite{2017pointer, 2017GRU}.
To assess parallel decoding (PD)'s efficiency through decoding-head cohorts, we compare DEGR(w/o PD) and DEGR.
The parameter count of DEGR's generator is slightly above DEGR(w/o PD) because of multiple FFN cohorts in G-Decoder.
% To compare complexity and time efficiency, we conduct all experiments on NVIDIA A100 GPU.
To ensure a fair comparison of time efficiency, all experiments are conducted on identical GPUs.
% Compared with DEGR (w/o PD), DEGR performs parallel decoding through decoding head queues, which improves time efficiency without reducing performance, as shown in Table \ref{tab:ablation}, \ref{tab:complexity}.
Compared with DEGR (w/o PD), DEGR performs parallel decoding and improves time efficiency, as shown in Table\ref{tab:complexity}.
During online serving, TP99 increases by 3.2ms, which is acceptable in recommendation systems.
The increase can be further optimized through engineering optimization, such as operator merging.

\begin{table}
  \caption{Summary of Complexity and Time Efficiency}

  \label{tab:complexity}
  \begin{tabular}{p{1.89cm}|p{1.0cm}|p{0.9cm}|p{0.87cm}|p{0.87cm}|p{0.4cm}}
    \toprule
    {\multirow{2}{*}{Method}} & \multicolumn{2}{c|}{Param Count} & \multicolumn{2}{c|}{Batch Time}&\multicolumn{1}{c}{Online}   \\
    {} & G & E / R  & G & E / R & {TP99} \\
    \midrule
    PRM     & 8.08kw & - & 0.925s & - & - \\
    GRN     & 24.84kw & 8.41kw  & 1.444s & 0.726s & - \\
    CMR     & 24.72kw & 8.41kw  & 1.448s & 0.715s& - \\
    NAR4Rec & 11.39kw & 8.41kw & 1.034s  & 0.721s & - \\
    GReF & 13.60kw & - & 1.013s  & - & - \\
    \hline
    DEGR(w/o PD) & 11.71kw & 8.41kw & 1.738s  & 0.732s & +7.8ms \\
    DEGR  & 13.60kw & 8.41kw & 1.388s & 0.732s & +3.2ms\\
    \bottomrule
  \end{tabular}
  \begin{tablenotes}
     \item[1] TP99 is the 99th percentile of online latency.
     % Here, we compare our method with baseline PRM.
   \end{tablenotes}
\end{table}

\section{CONCLUSION}
In this paper, we propose a Dual Exploration-Driven Generative Re-Ranking (DEGR) method to model exploratory value.
The dual exploration, hybrid supervised-reinforcement exploration and optimization paradigm, guided by an exploratory reward model that adaptively balances immediate and exploratory value, finally helps to build an end-to-end sequence optimizer as an adaptive cross-request contextual bridge.
DEGR outperforms baselines, contributing up to 1.22\% UCTR and 0.20\% PV in the JD recommendation system.
To better guide the generator's learning, we aim to build a more precise reward model by incorporating richer future user-behavior signals and exploring scaling laws for model parameters.

\bibliographystyle{ACM-Reference-Format}
\bibliography{sample-base}

\appendix

\balance
\section{Appendix}
\subsection{AR-ORPO Loss Pseudocode}
\definecolor{commentcolor}{RGB}{0,128,0} % 绿色
\renewcommand{\algorithmiccomment}[1]{\hfill\textcolor{commentcolor}{#1}}
\algrenewcommand\alglinenumber[1]{\footnotesize\ttfamily#1:}
Here, concerning code privacy, we provide a pseudocode of our proposed AR-ORPO loss calculation.
\begin{algorithm}
\caption{A Tensorflow-style Pseudocode of our proposed AR-ORPO Loss.}
\label{alg:example}
\begin{algorithmic}[1]
    % \Function{Cal\_Dynamic\_ORPO\_Loss}{ }
\State \textbf{Input}: seq\_reward: $[bs, |G|]$, prob\_multi: $[bs, |G|]$, quantiles $S$, temperature parameter $\tau$, batch size $bs$, sampled sequences $G$.

\State min\_idx $\gets$ argmin(seq\_reward, axis=1)
\State max\_idx $\gets$ argmax(seq\_reward, axis=1)
\State mid\_k $\gets |G| // (S - 1) $
\State \_, mid\_idx $\gets$ top\_k(seq\_reward, k=mid\_k $\cdot$ (S - 2))

\State key\_points $\gets$ [max\_idx] 
\State prob\_points$\gets$ [gather(prob\_multi, max\_idx)]
\State reward\_points $\gets$ [gather(seq\_reward, max\_idx)] 

\State for j in range(1, S-1): \Comment{\# Collect data at each quantile}
\State \ \ \ \ cur\_idx $\gets$ mid\_idx[:, mid\_k $\cdot$ j - 1]
\State \ \ \ \ key\_points.append(cur\_idx)
\State \ \ \ \ prob\_points.append(gather(prob\_multi, cur\_idx))
\State \ \ \ \ reward\_points.append(gather(seq\_reward, cur\_idx))

\State key\_points.append(min\_idx)
\State prob\_points.append(gather(prob\_multi, min\_idx))
\State reward\_points.append(gather(seq\_reward, min\_idx))

\State odds\_points $\gets$ []
\State for prob in prob\_points: \Comment{\# Caculate odds}
\State \ \ \ \ odds $\gets$ max(exp(prob) / (1 - exp(prob)), $1e-8$)
\State \ \ \ \ odds\_points.append(odds)

\State  weighted\_rewards $\gets$ [ exp( (r - reward\_points[0] ) /$\tau$) for r in reward\_points ] \Comment{\#Calculate weighted rewards}
\State  sum\_weighted\_rewards $\gets$ sum(weighted\_rewards)
    
\State loss\_components $\gets$ []
\State for j in range(S-1):
\State \ \ \ \ odds\_ratio $\gets$ log(odds\_points[j]) / sum(odds\_points[j+1:]) \Comment{\#Calculate odds ratio of the current and inferior quantiles}
\State \ \ \ \ sig\_ratio $\gets$ sigmoid(log\_odds\_ratio)
\State \ \ \ \ loss\_component $\gets$ log(sig\_ratio)
\State \ \ \ \ weight $\gets$ weighted\_rewards[j] / sum\_weighted\_rewards \Comment{\# Compute the soft weights from weighted rewards} 
\State \ \ \ \ loss\_components.append(weight $\cdot$ loss\_component)

\State loss\_A $\gets$ sum(loss\_components)
\State \textbf{return} loss\_A
    % \EndFunction
\end{algorithmic}
\end{algorithm}

\subsection{AR-ORPO Gradient Derivation}
As described in Section4.2.2, our adaptive reward-weighted ORPO (AR-ORPO) loss is defined as:
\[
    L_{A} = \sum_{i=1}^{S-1} \underbrace{\frac{e^{R_{\theta}(\tau_i, \ q) / t }}{ \sum_{j=1}^{K} e^{R_{\theta}(\tau_j, \ q) / t}  }}_{soft \ \ weights} \cdot \log \sigma \left( \log \frac{odds(\tau_i \mid q)}{\sum_{j=i+1}^{S} odds(\tau_j \mid q)} \right)
\]
Suppose soft weights $w_i=\frac{e^{R_{\theta}(\tau_i, \ q) / t }}{ \sum_{j=1}^{K} e^{R_{\theta}(\tau_j, \ q) / t}  }$, and $g(\tau_i,\tau_j)=\frac{\mathrm{odds}(\tau_i \mid q)}{\sum_{j=i+1}^{S} \mathrm{odds}(\tau_j \mid q)}$.

\[
\begin{aligned}
\nabla_\theta \mathcal{L}_{A}
&= \sum_{i=1}^{S-1} w_i \nabla_\theta \log \sigma \left( \log \frac{\mathrm{odds}(\tau_i \mid q)}{\sum_{j=i+1}^{S} \mathrm{odds}(\tau_j \mid q)} \right) \\
&= \sum_{i=1}^{S-1} w_i \sigma(-\log g(\tau_i,\tau_j)) \cdot \nabla_\theta \log g(\tau_i,\tau_j) \\
&= \sum_{i=1}^{S-1} w_i (1 + g(\tau_i,\tau_j))^{-1} \cdot \nabla_\theta \log g(\tau_i,\tau_j)
\end{aligned}
\]

\[
\begin{aligned}
\nabla_\theta \log g(\tau_i,\tau_j) 
&= \nabla_\theta \log \frac{\mathrm{odds}(\tau_i \mid q)}{\sum_{j=i+1}^{S} \mathrm{odds}(\tau_j \mid q)} \\
&= \nabla_\theta \log \mathrm{odds}(\tau_i \mid q) - \nabla_\theta \log ({\sum_{j=i+1}^{S} \mathrm{odds}(\tau_j \mid q)}) \\
&= \nabla_\theta \log \mathrm{odds}(\tau_i \mid q) - \frac{\sum_{j=i+1}^{S} \nabla_\theta \mathrm{odds}(\tau_j \mid q)}{\sum_{k=i+1}^{S}\mathrm{odds}(\tau_k \mid q)} 
\end{aligned}
\]
We calculate $\nabla_\theta \log \mathrm{odds}(\tau_i \mid q)$ and $\nabla_\theta \mathrm{odds}(\tau_j \mid q)$, respectively.
\begin{flalign*}
\nabla_\theta \log \mathrm{odds}(\tau_i \mid q) 
&= \nabla_\theta \log P(\tau_i \mid q) - \nabla_\theta \log (1-P(\tau_i \mid q)) && \\
&= \frac{1}{1-P(\tau_i \mid q)} \cdot \nabla_\theta \log P(\tau_i \mid q) &&
\end{flalign*} 

\begin{flalign*}
\nabla_\theta \mathrm{odds}(\tau_j \mid q)
&= \frac{\nabla_\theta P(\tau_j \mid q)}{(1-P(\tau_j \mid q))^2} &&\\
&= \frac{P(\tau_j \mid q) \cdot \nabla_\theta \log P(\tau_j \mid q)}{(1-P(\tau_j \mid q))^2} &&\\
&= \frac{\mathrm{odds}(\tau_j \mid q)}{1-P(\tau_j \mid q)} \cdot \nabla_\theta \log P(\tau_j \mid q)  &&
\end{flalign*} 
\[
\begin{aligned}
&\nabla_\theta \log g(\tau_i,\tau_j) 
= \nabla_\theta \log \mathrm{odds}(\tau_i \mid q) - \frac{\sum_{j=i+1}^{S} \nabla_\theta \mathrm{odds}(\tau_j \mid q)}{\sum_{k=i+1}^{S}\mathrm{odds}(\tau_k \mid q)} \\
&= \frac{\nabla_\theta \log P(\tau_i \mid q)}{1-P(\tau_i \mid q)} - \sum_{j=i+1}^{S}\frac{  \mathrm{odds}(\tau_j \mid q)}{\sum_{k=i+1}^{S}\mathrm{odds}(\tau_k \mid q)} \cdot \frac{\nabla_\theta \log P(\tau_j \mid q)}{1-P(\tau_j \mid q)}
\end{aligned}
\]
The final gradient of AR-ORPO is defined as:
\begin{flalign*}
&\nabla_\theta {L}_{\text{A}} = \sum_{i=1}^{S-1} \delta(d) \cdot h(d) , \ d=(\tau_i, \tau_j) \sim D  \\
&\delta(d) = w_i \cdot (1+\frac{\mathrm{odds}(\tau_i \mid q)}{\sum_{j=i+1}^{S} \mathrm{odds}(\tau_j \mid q)})^{-1}   \\
&h(d) = \frac{\nabla_\theta \log P(\tau_i \mid q)}{1-P(\tau_i \mid q)} - \sum_{j=i+1}^{S} \underbrace{ \frac{  \mathrm{odds}(\tau_j \mid q)}{\sum_{k=i+1}^{S}\mathrm{odds}(\tau_k \mid q)} }_{Odds \ normalized \ weight} \cdot \frac{\nabla_\theta \log P(\tau_j \mid q)}{1-P(\tau_j \mid q)}
\end{flalign*} 

Compared with the gradient of ORPO:
\begin{flalign*}
&\nabla_\theta {L}_{ORPO} = \delta'(d) \cdot h'(d) \\
&\delta'(d) = \left( 1 + \frac{\mathbf{odds}_\theta P(\tau_w|x)}{\mathbf{odds}_\theta P(\tau_l|x)} \right)^{-1} \\
&h'(d) = \frac{\nabla_\theta \log P_\theta(\tau_w|x)}{1 - P_\theta(\tau_w|x)} - \frac{\nabla_\theta \log P_\theta(\tau_l|x)}{1 - P_\theta(\tau_l|x)} 
\end{flalign*}

Assuming i.i.d. logit noise $\epsilon \sim \mathcal{N}(0, \sigma^2)$, AR-ORPO's variance satisfies $\text{Var}(\nabla_{L_A})\approx\frac{1}{S-1} \text{Var}(\nabla_{L_{ORPO}})$. Increasing S improves the determinism of the gradient direction. Soft weights $w_i$ transform models from static alignment to reward-aware importance weighting, effectively suppressing low-quality gradient noise.

Although we set $S=3$ and $t=0.2$ empirically (Fig.\ref{fig:orpo_trends}), they achieve a favorable trade-off between information gain and numerical stability. As S increases, discriminative signals tend to be canceled out in high-dimensional spaces, and easy negatives dilute the gradient weights of hard negatives. t in soft weights modulates curvature's steepness. As $t\to\infty$, $w_i$ tends to be uniform, which reduces variance but discards discriminative rewards.

\end{document}